\documentclass[10pt,conference]{IEEEtran}
\usepackage{pdfpages}
\usepackage{graphicx}
\usepackage[caption=false]{subfig}
\usepackage{color}
\usepackage{algorithmic}
\usepackage[ruled,vlined,longend,linesnumbered]{algorithm2e}
\usepackage{multirow}
\usepackage{tabularx,ragged2e}
\newcolumntype{L}{>{\RaggedRight\arraybackslash}X}
\usepackage[numbers,sort&compress,square]{natbib}

\usepackage[hyphens]{url}
\usepackage{hyperref}
\usepackage[hyphenbreaks]{breakurl} %for url break

\begin{document}

\title{Human-Centric Issues in eHealth App Development~and Usage: A Preliminary Assessment}
\author{\IEEEauthorblockN{Md. Shamsujjoha$^\dag$, John Grundy$^\dag$, Li Li$^\dag$, Hourieh Khalajzadeh$^\dag$, and Qinghua Lu$^\ddag$}
\IEEEauthorblockA{$^\dag$ Department of Software Systems and Cybersecurity, Faculty of Information Technology, Monash University, Australia 
\\
$^\ddag$ Data61, CSIRO,  Australia.\\
Email: \{md.shamsujjoha, john.grundy, li.li, hourieh.khalajzadeh\}@monash.edu, qinghua.lu@data61.csiro.au}
}
\maketitle

\begin{abstract}
Health-related mobile applications are known as eHealth apps. These apps make people more aware of their health, help during critical situations, provide home-based disease management, and monitor/support personalized care through sensing/interaction. eHealth app usage is rapidly increasing with a large number of new apps being developed. Unfortunately, many eHealth apps do not successfully adopt Human-Centric Issues (HCI) in the app development process and its deployment stages, leading them to become ineffective and not inclusive of diverse end-users. This paper provides an initial assessment of key human factors related to eHealth apps by literature review, existing guidelines analysis, and user studies. Preliminary results suggest that \textit{Usability}, \textit{Accessibility}, \textit{Reliability}, \textit{Versatility}, and \textit{User Experience} are essential HCIs for eHealth apps, and need further attention from  researchers and practitioners. Therefore, outcomes of this research will look to amend support for users, developers, and stakeholders of eHealth apps in the form of improved actionable guidelines, best practice examples, and evaluation techniques. The research also aims to trial the proposed solutions on real-world projects. \end{abstract}
%  No change required in abstract as per reviewers

%\hyphenation{op-tical net-works semi-conduc-tor} % correct bad hyphenation here


\section{Introduction}
\label{Introduction}
At the time of writing this paper, more than 43\% of the world's population had a smartphone, and an additional 13\% had a mobile device such as cell phones, tablets, and cellular-enabled~devices~\cite{Confirmation_1_John,confirmation_2_Carroll,confirmation_3_Statistica}. 
Felgoise et al.~\cite{confirmation_4_ironsource} reports that (i) mobile phone usage for different purposes is increasing at 7.71\% per year, and (ii) more than 205 billion mobile apps were downloaded from the cloud app repositories in 2018. Among billions of such apps, eHealth apps have become extremely popular recently, as these apps help people take greater control over their health and live more healthier lives. Moreover, demand from health professionals to push monitoring, education, and care plan implementation gear up the eHealth application usages, consequently its development~\cite{eHealth_infulence}. 

Human factors have significantly impacted how eHealth apps are developed and used~\cite{Confirmation_1_John}. We define these \textbf{H}uman-\textbf{C}entric \textbf{I}ssues (HCI) as key characteristics of human users, and include things such as age, culture, gender, cognitive ability, emotions, language,  educational attainment, socio-economic status and personality. Incorporating HCIs into the eHealth app development process is challenging, e.g., addressing the need of end-users who are aged, have a wide range of physical and mental challenges, have diverse languages, cultures and socioeconomic backgrounds.

Consider an eHealth app `SleepNea' that helps clinicians use to continuously monitor a sleep apnea patient's breathing and oxygen from a remote location.  Data needs to be updated continuously to provide real-time information. This information can be updated every moment or batch uploaded after an interval. Dealing with the sensor data and handling network issues are technical domain concerns. However, the design and working procedure of the app must also deeply employ and appreciate the human aspects, i.e., relative physical and mental challenges, for example, app usages, data exchange through sensors, and use of the extra device should not affect the day to day life of the patients, their families and friends, clinicians and community workers. It also includes the \textit{Technology Proficiency} and \textit{Acceptance} by the users who are likely to have different cultures, languages, and ages than the app developers and software engineers~\cite{confirmation_11_JohnGurndy}. 

The development of `SleepNea' app should factor in the \textit{Emotional} – both positive and negative – reactions to the app, e.g., up to date feedback and suggestion is potentially positive but being continuously monitored potentially negative. The \textit{Accessibility} of the solutions for people with allergies, physical tremors, poor eyesight, or wheel-chair bound is cognitive decline. The \textit{Usability} of this app for a group of people should also address the varied needs, incorporating the use of sensors and modified smartphone interface, accommodating different ages, genders, cultures and languages of users, including appropriate use of text, colors and symbols. This is particularly important as one-quarter of the elderly in Australia are non-native English speakers and the majority are women, but by far, the majority of software developers are 20-something-year-old English-speaking men, the same as in the United States and dominant English-speaking countries. 
Within this, \textit{Personality Differences} may be very important, e.g., those who want flexible dialogue with doctors compared to those needing directive suggestions from the app itself. Failure to incorporate such HCIs can result in a mobile app that is unsuitable for whom it is designed for by introducing confusing, possibly unsettling and invasive, and even potentially dangerous technology~\cite{HCI_1_John_Emotion}. 

We want to develop a more integrated approach for eHealth app development and usage that addresses such human-centric issues of their users. We also want to find out how different HCIs are being addressed by developers now, which ones are missing/poorly handled, and which ones are the most important for different end-user groups. Our research also looks at developing ways to improve support for users, developers, and other stakeholders of eHealth apps in the form of improved actionable guidelines,  work-flow framework, tools, best practice examples, and evaluation techniques. These proposed solutions will also be trialed with different stakeholders of eHealth apps through an example eHealth development project and its produced apps. To do these, we are reviewing a set of existing eHealth apps, relevant literature, current development procedures, evaluation guidelines, and conducting surveys and interviews among eHealth apps end-users, stakeholders and developers. To this end, this paper presents (i) an analysis of gaps and recommendations about key HCIs for eHealth apps, (ii) some preliminary results from our pilot user studies, and (iii) a discussion of current and planned future research.

\section{Research Methodology}
\label{Sec_2_Reseach_methodology}
This section introduces our key \textbf{R}esearch \textbf{Q}uestions (RQs). We then discuss the current progress and preliminary results for answering these research questions in the following sections, including the relative threats and next research plans. 

\subsection{Research Questions (RQs)}
\label{vissionRQs}
\begin{description}
\item[RQ1:] Which human-centric issues in eHealth apps are most important for different user groups, e.g., patients, health practitioners, researchers, developers, and associate members?

\item[RQ2:] How do stakeholders identify human-centric issues in the eHealth app domain, and what should be integrated for more effective usage, and consequently, its development?

\end{description}

We are conducting several related research activities to answer the above research questions, summarized below: 

\begin{itemize}
\item We reviewed a number of authenticated and widely used existing guidelines that include HCIs and health applications to analyze the first RQ. A summary of our analysis results are discussed in section Sec.~\ref{Sec_3_Existing Guidelines}. 

\item  We analyzed a set of eHealth apps based on existing guidelines and using our own experience while using these apps in different scenarios. These analysis results are shared for the research community in~\cite{MARS-and-Dictionary-SANER2021-GITHub}. 

\item We collected 4-major medical dictionaries and extracted 99,866 keywords to identify the development patterns, including HCIs which is also shared in~\cite{MARS-and-Dictionary-SANER2021-GITHub}. 

\item We are conducting two different user studies focusing on developers and end-users/stakeholders of the eHealth apps to find more detailed insight for answering RQ2. \end{itemize}

\begin{table*}[htbp]
\centering
\caption{Essential Human-Centric Issues for eHealth Apps: Current Practice, Gaps and Recommendation}
\label{Vission_Table_1_Gaps_Recom}
\begin{tabular}{|c|p{1.75in}|p{1.4in}|p{2.6in}|}
  \hline
      \textbf{HCIs} &\textbf{Investigated for eHealth Apps} & \textbf{Major Gaps} & \textbf{Recommendations} \\ \hline
       Usability 
       & Taps, Mode (landscape vs portrait), Platforms, OS versions, Resolution, Data fill-up, Responsiveness, Content. 
       & Aesthetic and minimalist design, Less user control, Inconsistent and low standards for content. 
       & Design, develop, and evaluate eHealth apps for diverse users using usability focused methodologies.\\ \hline
      
      Accessibility 
      & Text contrast, Alternate text vs images, Links, Navigation, Form, Labels, Table, Time-outs, Sitemap.  
      & Display variance, Undermine users, No/low repair support and assistive technologies.
      & Emphasized the input modalities and specific need of target audiences during design, development and deployment (app usages).  \\ \hline
      
      Reliability  
      & Security, Privacy, Dependability, Robustness,  and Trustworthiness.  
      & Information processing, Synchronization, Platform, Independence. 
      & Identify potential failures point beforehand to perform equally well even when unexpected events occur through a secured mechanism.  \\ \hline
      
      Versatility 
      &  App information, Users issues e.g., age, language, technology proficiency.  
      & Adaptive interfaces, User diversity, Cognitive aspects
      & Provide variety of health information / communication services at different levels in the user community domain.\\ \hline
      
      User Experience 
      & Presentation, Functionality, Ease of use, Performance. 
      & Guidance, Problem diagnosis (usages), Assistance to use
      & Include resource-aware mechanisms that incur negligible overhead, are assistive for service and interactive.\\ \hline

\end{tabular}
\end{table*}

\section{Important Human Centric Issues for eHealth Apps: Evaluation Results and Analysis for RQ1}
\label{Sec_3_Existing Guidelines}
Eshet et al.~\cite{Mobile_HCI_survey} reviewed 79 research papers that investigate HCIs for mobile app development until 2012. Since then, best practices have been revised in user-developer communities. We evaluated more recent literature in this domain to answer RQ1, and these review and analysis results are summarized below: \\
\\
\textbf{Mobile App Rating Scale (MARS)~\cite{HCI_Review_SANR2-MARS}:} MARS is a simple and reliable tool for measuring the quality of mobile health apps. It provides a checklist for high-quality eHealth app design and development. Firstly, it evaluates apps through two sources -app targets, and -theoretical strategies. Then, it evaluates engagement, functionality, aesthetics, information and recommendation for app quality measurement. 
\\
\\
\textbf{Health Apps Assessment Guidance~\cite{HCI_Review_SANR1-NZ}:} This guideline was prepared by the New Zealand Ministry of Health wings. The guideline focuses on two key areas - guidance for clinicians and consumers that include key points to consider for selecting an appropriate health app, and evaluate the effectiveness of a health app, and -guidance for app developers that include key points to consider before deciding to develop a new health app.
\\
\\
\textbf{Health Navigator Guidance for App Developers~\cite{HCI_Review_SANR3-nz}:} It comprises an integrated list of resources in eHealth app development, co-design and information processing for \textit{Usability} and \textit{Health Literacy}. 
\\
\\
\textbf{Digital Health App Development Standards~\cite{HCI_Review_SANR4-protocol}:} This article is a systematic review protocol that provides an overview of the current standards, frameworks,  guidelines, and best practices for eHealth app development. 	
\\
\\
\textbf{Award Winning Health App~\cite{HCI_Review_SANR5-au}:}	This study tried to find gaps in information sharing during eHealth apps usages, particularly for patients with complex conditions who visit multiple health service providers. It suggests treating the user as a person and recognizing what matters most for them, e.g., `a patient is a person, not just a collection of symptoms, and an eHealth app (data) should assist clinicians in treating the whole person.' It also recommends best practices for time, data, and record management in the app to incorporate HCIs.
\\
\\
\textbf{W3C/WAI Guidelines~\cite{HCI_Review_SANR7-w3c}:}  W3C/WAI guideline defines the \textit{Accessibility} and summarizes its applicability for mobile platform. Main four discussed issues are - \textit{Perceivable} i.e., screen size, information presentation, content and touch control, magnification; - \textit{Operable} i.e., keyboard control, touch target size and spacing,  gestures, device manipulation; - \textit{Understandable} i.e., orientation, layout, page positioning, grouping; - \textit{Robust} i.e., virtual keyboard, data entry method, platform properties. The guideline also exemplifies each issue with evidence for mobile devices.
\\ 
\\
\textbf{mHealth Interventions for Vulnerable Populations~\cite{HCI_Review_SANR11-Conf}:} The impact of mHealth (eHealth) tools/frameworks for low-socioeconomic, racial, and ethnic minority groups are reviewed in \cite{HCI_Review_SANR11-Conf}. Initially, authors illustrate  \textit{Usability} synthesis across different eHealth projects focused on users' health equity and then analyze the user experience findings (\textit{Feasibility}). It is also shown that the Health Belief Model, Trans Theoretical Model of behavior change, Social Cognitive and Goal-setting theories are most examined in this domain. 
\\
\textbf{eHealth App Development for Elderly User~\cite{HCI_Review_SANR8elsevier,HCI_Review_SANR-12Journal}:} Issues related to vision (icon, font size and type), hearing, and spatial coordination in a health app for older people are discussed in~\cite{HCI_Review_SANR8elsevier}. This article also suggests how to avoid such \textit{Usability} and \textit{Accessibility} challenges in future health apps. In \cite{HCI_Review_SANR-12Journal}, a set of eHealth apps and corresponding guidelines are analyzed. The authors then suggest a compact checklist for health app development and usages, where older adults are the primary target audience.
\\
\\
% \textbf{Mobile Web Accessibility~\cite{HCI_Review_SANR10-Conf}:} This paper examined  \textit{Accessibility} issues for web content developers, separated as mobile and non-mobile contexts.
% \\
% \\
\textbf{Factsheet~\cite{HCI_Review_SANR15-vic}:} A fact-sheet was proposed by the Victorian health authority in~\cite{HCI_Review_SANR15-vic}. It includes three detailed steps for helping the medical practitioner for assessing the quality of a healthy living app and advise their patients on what to look for.
\\
\\
\textbf{World Health Organization (WHO) guideline~\cite{HCI_Review_SANR14-who}:} The key aim of the WHO guideline is to present recommendations for health system improvements by evaluating current evidence, especially on emerging digital and mobile health interventions that are contributing. The criterion included are based on an assessment of - \textit{Benefits}, \textit{Harms}, \textit{Acceptability}, \textit{Feasibility}, \textit{Resource usages}, and \textit{Equity considerations}. This guideline also represents a subset of prioritized digital health interventions that are accessible via mobile devices. 
\\
\\
\textbf{Preliminary Results:} Our preliminary results are summarized in Table~\ref{Vission_Table_1_Gaps_Recom}, where we define mostly examined HCIs for eHealth apps with gaps and recommendations, but by no means, these are all of the key issues we are interested in, have worked on, or are working on currently. This table also shows that further investigation is needed for appropriately incorporating HCIs in eHealth apps, especially for diverse users and stakeholders. 

\section{Identify Stakeholder Concerns for eHealth Apps:  Evaluation Results and Analysis for RQ2}
\label{Sec_4_user_study}
In RQ2, our primary aim is to find a big picture view of current practices, challenges, and approaches being used to address the HCIs in eHealth apps and fundamental future needs. For this purpose, two different user studies are designed. The first user study will look at the current mobile app developer engagement to support HCIs in eHealth apps. In contrast, the second one will look for the end-user/stakeholder related issues. % We already obtained full Human Subject Ethics Committee approval for this study. 
The specified target participants are:

\begin{itemize}
\item \textbf{Mobile app developer:} We are particularly interested in surveying the mobile app developers experienced in different domains such as front-end and back-end development, data processing, quality assurance (QA), team lead, project management and so on.  

\item \textbf{eHealth app stakeholders and users:} We are interested in different types of end-users and stakeholders of the eHealth apps e.g., users with physical/mental challenges, users with different age groups, non-English speakers, users with cultural diversity, low socio-economic status, low access to technology, and so on.  
\end{itemize}
\begin{table*}[htbp]
\caption{Preliminary Results of the Surveys (Summarized)}
\label{User_study_Prili_results}
\begin{tabular}
{|p{.15 in}|
>{\raggedright\arraybackslash}p{0.9 in}|
>{\raggedright\arraybackslash}p{1.25 in}|
>{\raggedright\arraybackslash}p{1.25 in}|
>{\raggedright\arraybackslash}p{1.25 in}|
>{\raggedright\arraybackslash}p{1.25 in}|
>{\raggedright\arraybackslash}p{1.5 in}}
\cline{1-6}

\multicolumn{2}{|p{.3 in}|}{\multirow{2}{*}{\textbf{Criterion}}} & \multicolumn{2}{c|}{\textbf{Pilot Study-1}} &
  \multicolumn{2}{c|}{\textbf{Pilot Study-2}} &
   \\ \cline{3-6}
\multicolumn{2}{|c|}{} &
  \multicolumn{1}{c|}{\textbf{Developer   Survey}} &
  \multicolumn{1}{c|}{\textbf{End-User   Survey}} &
  \multicolumn{1}{c|}{\textbf{Developer   Survey}} &
  \multicolumn{1}{c|}{\textbf{End-User   Survey}} &
   \\ \cline{1-6}
\multirow{5}{*}{\rotatebox[origin=c]{90}{\textbf{Ethnographic Information\,\,\,}}} &
  Countries &
  Australia~(20\%), Bangladesh~(60\%), Canada~(20\%) &
  Australia~(16.67\%), Bangladesh~(50\%), Japan~(33.33\%) &
  Australia~(25\%), Bangladesh~(50\%), Canada~(25\%) &
  Australia~(40\%), Bangladesh~(20\%), Italy~(20\%), UK~(20\%) &
   \\ \cline{2-6}
 &
  Age Groups &
  21-30~(20\%), 31-40~(80\%) &
  21-30~(66.67\%), 31-40~(33.33\%) &
  21-30~(50\%), 31-40~(25\%), 51-60~(25\%) &
  21-30~(40\%), 31-40~(40\%), 61-70~(20\%) &
   \\ \cline{2-6}
 &
  Gender &
  Male~(60\%), Female~(40\%) &
  Male~(66.67\%), Female~(16.67\%),~N/A~(16.67\%) &
  Male~(75\%), Female~(25\%) &
  Male~(40\%),~Female~(40\%), Non-binary~(20\%) &
   \\ \cline{2-6}
 &
  Qualification &
  BSc~(100\%) &
  MBBS~(33.33\%),~BSc (33.33\%), MS~(33.33\%) &
  Diploma~(25\%), BSc~(75\%) &
  Secondary~(20\%),~MBBS (20\%),~BA~(20\%),~PhD~(40\%) &
   \\ \cline{2-6}
 &
  Experience   in App Development/Usage &
  4.4   year on average (development) &
  6.17   year on average (usages) &
  6.0    year  on average (development) &
  7.0   year on average (usages) &
   \\ \cline{1-6}
\multirow{3}{*}{\rotatebox[origin=l]{90}{\textbf{Views: HCIs and eHealth app}\,\,\,}} &
  HCIs for eHealth Apps &
  Absolutely   Essential: 70\%, Important: 30\% &
  Absolutely   Essential: 73.33\%, Important: 10\%, N/A: 16.67\% &
  Absolutely   Essential: 78.57\%, Important: 21.42\% &
  Absolutely   Essential: 57.14\%, Important: 14.28\%, N/A: 14.28\% &
   \\ \cline{2-6}
 &
  Critical HCIs&
  \multicolumn{4}{l|}{Reliability, Accessibility, Usability, Versatility (app and user), User Experience (in that order and merged)} 
 
   \\ \cline{2-6}
 & Themes (Open Ended Questions, Comments and Suggestion) 
 
 & \textit{Accessibility} and \textit{Usability} are most important HCIs, young users are more adaptive to app usages than the older adult, app components should follow the current practice, user-feedback is crucial. 
  
 & Data privacy and security are of significant concerns, precise and reliable health information should be advised during app usages, app presentation largely motivate users. 
 
 & Emphasize \textit{Accessibility}, tutorials smoothen app usages for understanding app features/functionalities, improved prototyping schemes are needed for future health apps. 
 
 & Should be authorized (or check with hospital/clinic) before suggesting health advice (especially in critical conditions), alerts/notification help users, more automated emergency services and payment is needed. 
   \\ \cline{1-6}
\end{tabular}
\end{table*}

\textbf{\\Preliminary Results:}
%\label{PL_Results}
A pilot study using our surveys was conducted from August to September 2020 among relevant populations. Some preliminary results are summarized in Table~\ref{User_study_Prili_results}. Currently, we are running these surveys in a full data gathering phase to get a much wider pool of responses. In the pilot runs, respondents over six different countries considered the HCIs in eHealth apps are essential, where \textit{Reliability}, \textit{Accessibility}, and \textit{Usability} issues need further attention. The respondents were particularly satisfied with current apps versatility but are less satisfied with adaptive service, user interface and security (mainly app data). Overall, most participants suggest that the eHealth app should be trustworthy and, if possible, authorized. 

In the next section, we briefly present our plans to solve the identified shortcoming. However, discussion such as \textit{how the user study is being conducted}? \textit{What are the questions?}  are beyond this paper's scope. 

\section{The Next Step}
\label{Sec_5_Next Step}
We plan to selectively interview the two key participant groups to find out more detail on HCIs that emphasize current eHealth app usages and development. In other words, we want to better understand the challenges to address these issues from a human-centric perspective, get feedback on key deficiencies with the current app they try and use, and triangulate with the above findings. The idea is to enhance the broad picture obtained from the survey to drill down to more specific information. Then, we will summarize the crucial findings from these user studies. We will also evidence why the identified gaps are considerable problems and why these need to be resolved. In the third step, we plan to exemplify how we can proficiently resolve these gaps to improve the produced eHealth apps, for example, how some of the existing HCIs such as disability- or accessibility- related issues can be addressed in the current environment or future protocol(s) design. Finally, we will trial the proposed solutions with the different stakeholders of eHealth apps in the form of feedback, ultimately on example projects and the generated apps, where we will examine the following criterion: 
\begin{itemize}
\item \textbf{Usability evaluation:} Completion rate, task time,  task-level satisfaction, test-level satisfaction, errors, expectation, conversion, single usability metric.  

\item \textbf{Reliability evaluation:} Mean time between failure, preventive maintenance. 

\item \textbf{Scalability evaluation:}  Throughput, resource usage, cost, performance, and capacity. 

\item \textbf{Threats to validity:} Conformance vs accessibility-in-use. 

\item \textbf{Impacts on Stakeholders:} Satisfaction and benefits.

\item \textbf{Sensitivity measurement:} Reflection to changes. 

\item \textbf{Adequacy adaptation:} Distribution, complexity and computational demand. 
\end{itemize}

\subsection{Data Analysis and Outcomes}
\label{Analysis-outcome}
We will use quantitative analysis (mainly from survey responses) and qualitative analysis (interview questions and survey open answer questions) to sum up the following outcomes: 

\begin{itemize}
\item Identify the range of HCIs that need to be considered by the developers, users and stakeholders of eHealth apps.
    
\item Analyze data from the full-surveys and interviews quantitatively (thematic analysis) and quantitatively (descriptive and explore associations). 
    
\item  Identify which HCIs are the more important, difficult, challenging to take into account/meet during eHealth usages and development? 
    
\item Identify if there are any particularly difficult HCIs to address depending on different user groups, developers, organizational context, stakeholder concerns, and domain restrictions. 
    
\item Develop an initial set of definitions, terminologies, and examples of HCIs for eHealth app users, developers, and other audiences for its more effective usage, development, and deployment. 
    
\item Identify the “best practices”  examples through the mentioned actionable guidelines. 
\item Prepare an analysis method and framework to address the HCIs that may be valuable for wider communities to inform about.
    
\item Evaluate the proposed solutions using a trial project, generated apps, and based on interviews-surveys. 
\end{itemize}

\section{Discussion}
\label{Sec_6_Discussion}
We are researching new approaches to mobile eHealth app development and usage that consider human aspects in the existing schemes. The project also aims to find new ways to determine which HCIs are essential to include when designing health applications, and how software developers can do this to increase its effectiveness during usage. The study will also look at ways people currently use eHealth apps in terms of HCIs. We are particularly interested in getting insight into eHealth apps developed for “challenged” people, e.g., those with physical or mental challenges, ageing users, people from low socio-economic backgrounds, those whose use of English language may be limited, and other vulnerable end-users. We are also interested in how other human aspects such as personality type, IT proficiency, emotional reactions, cognitive approaches, culture, ethnicity, level of engagement, and many more influence the eHealth apps. To our best knowledge, this is a novel idea in literature. 
% %
% \subsection{Threads to Validity}
% \label{Risk}

However, this research largely depends on the quality feedback both from mobile app developers and eHealth app stakeholders. Existing studies show that adapting human-centric features for diverse users is challenging~\cite{HCI_1_John_Emotion,Mobile_HCI_survey}. The target domain `eHealth app' and `HCIs' for this research might create an additional obstacle to this process. We also face deficiencies to fully understand the right balance in target communities and target solutions. For example, supporting a patient with 'Sleep Apnea', the app-based solution should not affect users' daily life. However, continuous monitoring in the app is needed (discussed in Sec.~\ref{Introduction}). Then, fulfilling observational constraints while maintaining human-centric requirements might hinder the apps' (Sleep Apnea's) practical and useful usages.

\section{Summary and Future Work}
\label{conclusion}
This research investigates Human-Centric Issues in mobile eHealth apps. Our aim is to increase the effectiveness of eHealth app usage and development. To this end, we presented a preliminary assessment of two-primary research questions along with research methodologies. We also explained the next research plan with corresponding risks and how we plan to address these challenges to assist different target communities. 

We are currently investigating two additional questions to extend this research: (i) \textit{How can we combine important  Human-centric factors into the eHealth app development and its analysis, evaluation, and usage}?  (ii) \textit{Do the augmented actionable guidelines, framework, evaluation method, and best practice examples that we are looking will improve the eHealth apps produced}? For this purpose, we will analyze our user study results, prepare the mentioned outcomes, and trial the proposed solutions in the real world. We are currently in the second phase out of three phases of this research project. The third phase is planned to be completed by December 2022. After that, we plan to design training units and run workshops to validate our approach. 

\section*{Acknowledgements}
Shamsujjoha is supported by Monash International Tuition Scholarship, RTP Stipend, and CSIRO Data61 Top-up Scholarship for his Ph.D. study at Monash University, Australia. 
This work was also supported by the Australian Research Council (ARC) under a Laureate Fellowship project FL190100035, a Discovery Early Career Researcher Award (DECRA) project DE200100016, and a Discovery project DP200100020.

\end{document}